\documentclass[,nonblindrev]{informs3} 
\OneAndAHalfSpacedXI 

\usepackage{endnotes}
\usepackage{bm}
\let\footnote=\endnote
\let\enotesize=\normalsize
\def\notesname{Endnotes}%
\def\makeenmark{$^{\theenmark}$}
\def\enoteformat{\rightskip0pt\leftskip0pt\parindent=1.75em
  \leavevmode\llap{\theenmark.\enskip}}
\usepackage{enumerate}

\usepackage{natbib}
 \bibpunct[, ]{(}{)}{,}{a}{}{,}%
 \def\bibfont{\small}%
 \def\bibsep{\smallskipamount}%
 \def\bibhang{24pt}%
 \def\newblock{\ }%
\TheoremsNumberedThrough     
\ECRepeatTheorems

\EquationsNumberedThrough    

\renewcommand{\theARTICLETOP}{}
\usepackage{booktabs}
\usepackage{multirow}
\usepackage[caption=false]{subfig}

\usepackage{graphicx}

\begin{document}
\RUNAUTHOR{Wang et al.}

\RUNTITLE{Clinical and Non-clinical Effects on Surgery Duration: Statistical Modeling and Analysis}

\TITLE{Clinical and Non-clinical Effects on Surgery Duration: Statistical Modeling and Analysis}

\ARTICLEAUTHORS{%
	\AUTHOR{Jin Wang}
	\AFF{Department of Systems Engineering and Engineering Management, City University of Hong Kong, Hong Kong, China, \EMAIL{wangjin.devin@my.cityu.edu.hk}} 
	\AUTHOR{Javier Cabrera}
	 \AFF{Department of Statistics and Biostatistics, Rutgers University, New Jersey, USA, \EMAIL{xavier.cabrera@gmail.com}}
	\AUTHOR{Kwok-Leung Tsui}
	\AFF{Department of Systems Engineering and Engineering Management, City University of Hong Kong, Hong Kong, China, kltsui@cityu.edu.hk}
	\AUTHOR{Hainan Guo}
	\AFF{Research Institute of Business Analytics and Supply Chain Management, College of Management, Shenzhen University, Shenzhen, China, 518060, guohn403@126.com}
	\AUTHOR{Monique Bakker}
	\AFF{Department of Systems Engineering and Engineering Management, City University of Hong Kong, Hong Kong, China, mbakker2-c@my.cityu.edu.hk}	
	\AUTHOR{John B. Kostis}
	\AFF{Cardiovascular Institute, Rutgers Robert Wood Johnson Medical School, New Brunswick, NJ, kostis@rwjms.rutgers.edu}
} 

\ABSTRACT{%
 Surgery duration is usually used as an input of the operation room (OR) allocation and surgery scheduling problems. A good estimation of surgery duration benefits the operation planning in ORs. In contrast, we would like to investigate whether the allocation decisions in turn influence surgery duration. Using almost two years of data from a large hospital in China, we find evidence in support of our conjecture. Surgery duration decreases with the number of surgeries a surgeon performs in a day. Numerically, surgery duration will decrease by 10 minutes on average if a surgeon performs one more surgery. Furthermore, we find a non-linear relationship between surgery duration and the number of surgeries allocated to an OR. 
 Also, a surgery's duration is affected by its position in a sequence of surgeries performed by one surgeon. In addition, surgeons exhibit different patterns on the effects of surgery type and position. Since the findings are obtained from a particular data set, We do not claim the generalizability. Instead, the analysis in this paper provides insights into surgery duration study in ORs.
 
} 

\KEYWORDS{Healthcare; surgery duration; workload; surgery sequence; surgeon} 

\maketitle

\section{Introduction}
Amidst rising costs of operating rooms (ORs), hospitals strive to satisfy high surgery demand with increasingly limited budgets. Costs in ORs largely depend on the allocation and scheduling decisions \citep{Denton2007}. Also, in literature relevant to surgery allocation and scheduling problems (\textit{e.g.}, \cite{denton2003sequential} and \cite{Ho-YinMak2014appointment}), surgery duration is considered as an important input. Hence, estimation of surgery duration is critical to hospital management. Evidence suggests that surgery duration depends on clinical factors, \textit{e.g.,} the surgery type (details of the procedure), the patient's anamnesis \citep{Lowndes2016}, as well as the surgeon's experience \citep{Zheng2008}. However, non-clinical impacts have not been investigated to the same extent. In this paper, we would like to investigate the impacts on surgery duration from a broad perspective, \textit{i.e.}, the clinical and non-clinical factors as well as the interactions between them.

The term ``surgery duration'' in this paper is somewhat different from that in literature. We define it explicitly as follows. The whole surgery process can be divided into four parts. The first part is the interval from the time a patient enters the OR to the start of anaesthesia, which is labelled preparation period. The second part is the interval from start to the end of anaesthesia, which we label anaesthesia period. The third part is the interval from the start (\textit{i.e.}, the end of anaesthesia) to the end of the surgical procedure, which we label surgery period. The last part is the interval from the end of the surgical procedure to the time the patient leaves the OR, which we call wake-up period.  The term ``surgery duration'' throughout the paper means the length of the third period, \textit{i.e.}, that of the surgical procedure itself, since this time period is dominated by the surgeon and critical for surgical outcome quality.

In this paper, non-clinical factors refer to day-of-the-week, surgeon workload, the workload in an OR, and the surgery position in a sequence of surgeries. The workload of a surgeon means the number of surgeries a surgeon performs in a day. We find that surgery duration will increase with around 10 minutes if a surgeon performs one more surgery. The workload in an OR refers to the number of surgeries scheduled in an OR in a day. Interestingly, we find a non-linear relationship between workload in ORs and surgery duration. When there are no more than 4 surgeries in an OR, surgery duration decreases with around 10 minutes if one more surgery is allocated to the OR; and when there are more than 4 surgeries in an OR, surgery duration increases with around 5 minutes for each additional surgery. Surgery position is the position of a surgery in a sequence of surgeries a surgeon performs in a day. We find evidence of a relationship between its position and its duration. In other words, the duration of a surgery varies with its position in a sequence. Furthermore, we investigate the effects of the interactions among surgery type, position, and surgeon, by which we aim to uncover whether different surgeons exhibit different patterns on the effects of surgery type and position. Indeed, we find evidence suggesting that for certain surgery types, its position is linked to surgeons working faster (or slower). Note that we do not claim the generalizability of the findings in this paper, since they are obtained based on a particular data set. Instead, our paper provides insights into the surgery duration analysis in ORs. In particular, similar analysis can be repeated for any OR to find the clinic and non-clinical imparts on surgery duration. Note that the results in the paper have been shown in \cite{wang2017predicting}. In this paper, we present details of our data and the methodologies used to obtain the results.

Our work makes the following contributions to literature on the impacts on surgery duration:
\begin{itemize}
  \item[(1)] 
  Usually, researchers use surgery duration as an input to make surgery allocation and scheduling decisions. However, we find that allocation and scheduling decisions in turn influence surgery duration.
  \item[(2)] We investigate the effects of two kinds of workloads on surgery duration, \textit{i.e.}, surgeon workload and workload in ORs.
  \item[(3)] We propose a new method to find the interactions between predictors. That is, the regression tree is used to indicate the possible interactions. 
  \item[(4)] We find that the surgery position in a sequence of surgeries a surgeon performs in a day impacts the surgery duration.
   \item[(5)] We find that surgeon performance is related not only to surgery types but also to the surgery  position in a day. 
\end{itemize}
The remainder of this paper is organized as follows. Section 2 gives a review of literature. Section 3 develops hypotheses. Section 4 describes the data structure. In Section 5, we propose statistical models and illustrate the results. Section 6 presents concluding remarks along with some possible future work.

\section{Literature Review}
We review the literature relevant to the factors that impacts surgery duration, \textit{e.g.}, the day of the week, the surgeon performing the procedure, and the surgeon workload.

The weekday or weekend effects in healthcare have been investigated. \cite{cram2004effects} find that the mortality rate of patients admitted to hospitals on weekends is slightly higher than that of patients admitted on weekdays. \cite{kostis2007} conclude that patients with myocardial infarction admitted on weekends experience higher mortality and lower use of invasive cardiac procedures. 

The impacts of the workload have been studied for decades. In industry, the empirical study of \cite{Schultz1999} illustrates that employees tend to work faster when they face rising inventory and the work is standardized. \cite{Oliva2001} find that employees can ``cut corners'' by omitting tasks to reduce waiting time for incoming customers. In healthcare, plenty of researches have investigated the effects of occupancy in medical ward on the length of stay (LOS), readmission rate, and mortality rate in hospitals. Some papers conclude that high occupancy levels lead to decreased LOS, \textit{e.g.}, \cite{Anderson2011} and \cite{Kc2012msom}. This kind of speed-up may have consequences such as  high readmission rates\citep{Kc2012msom} and high mortality rates \citep{Kc2009}. It is worth noting that the workload in ORs discussed in this paper is different from that in medical wards. First, the workload or occupancy of medical wards in the above papers is represented by the number of patients. The workload is work needed simultaneously by the patients in the ward, while the workload of a surgeon in a day in this paper is the number of patients who need surgery sequentially. Hence the workload of a surgeon should be time-dependent. Specifically, a surgeon may experience workload pressure during surgeries earlier in the day, but not during the latter surgeries. Or it may be the other way around. Second, different from patients in medical wards, a surgery is often performed by one surgeon or a surgeon team. Hence, the surgery duration is often related to the surgeon performing the procedure.

In the literature, surgery duration has often been linked to the surgeon performing the procedure. In particular, it is influenced by the performing surgeon's level of experience \citep{Zheng2008}, or to that of the assistant surgeon \citep{MolinaPariente2015}, or to that of the surgical team as a whole \citep{Zheng2008, Eijkemans2010}. Team size has also been investigated in relation to surgery duration \citep{Cassera2009}. 

Some scholars found evidence that surgery duration is related to procedure start time. \cite{Peskun2012} found that surgery duration increases with later surgery start time, but the small differences found (7.1 minutes) were likely not clinically significant. For procedures that started after 5 PM, \cite{Cassera2009} found no significant increase in surgery duration. Different from the papers that focus on start time, in this paper, we study the relevant effects on the surgery duration from another perspective, i.e., the position of a surgery in a sequence of surgeries performed by a surgeon. It is also an aspect to illustrate whether surgery duration depends on the surgeon performing the procedure.

\section{Hypothesis Development}\label{sec:hypo}
Generally, it is believed that surgery duration depends on surgery type, and on the details of the surgical procedure, \textit{e.g.,} how many tumours need to be removed, how deep are they, etc. However, we suspect that some non-clinical factors influence surgery duration. In this section, we develop hypotheses which will be tested in section \ref{sec:ana}.
\subsection{Hypothesis 1: the Effects of Weekdays }
Literature has investigated the effects of weekdays on mortality, \textit{e.g.}, \cite{cram2004effects} and \cite{crowley2009influence}. It is concluded that the effects of weekends are often significant, \textit{e.g.,} the mortality rate of patients admitted during the weekend is higher. In this paper, we investigate whether effects of the weekdays influence surgery duration significantly. We formally present the hypothesis as follows.
\begin{itemize}
  \item [\textbf{Hypothesis 1:}] The effects of weekdays influence the mean of surgery duration.
\end{itemize}
The corresponding \textit{null hypothesis} is that the effects of weekdays do not affect the mean of surgery duration.

\subsection{Hypothesis 2: the Effects of Workload} \label{sec:hypo-2}
 We focus on two kinds of workload: surgeon workload and workload in ORs. A surgeon might perform a different number of surgeries in one day. We conjecture that surgery duration decreases with the number of surgeries performed by the surgeon in a day. More specifically, surgeons under high workload will accelerate their surgeries. On the other hand, the number of surgeries that scheduled in an OR varies day by day. Hence, we also conjecture that if many surgeries are scheduled in an OR, the workload pressure in the OR will also cause surgeons to accelerate their surgeries. We present these conjectures formally in the following hypothesis.  
\begin{itemize}
  \item [\textbf{Hypothesis 2:}] the mean of surgery duration decreases with (1) the number of surgeries a surgeon performs in a day, and (2) the number of surgeries scheduled in the OR.
\end{itemize}

The corresponding \textit{null hypothesis} is that the mean of surgery duration does not vary with the number of surgeries a surgeon preforms in a day, or with the number of surgeries scheduling in the OR.

\subsection{Hypothesis 3: the Effects of Orderings}
We also suggest that surgery duration is affected by the position of the surgery in the sequence of surgeries that a surgeon performs in a day. Specifically, the duration of prior surgeries will be shorter, while that of the later surgeries will increase. Our reasoning is twofold. Firstly, if a surgeon performs multiple surgeries in a day, he/she may experience higher workload pressure when performing the earlier surgeries, knowing that there are still a number of surgeries to come. Hence, the surgeon will accelerate his/her work and the surgery duration will decrease. For latter surgeries we suppose the opposite. Secondly, if a surgery is one of the first procedures for a surgeon that day, it is usually one of the first surgeries in the OR in which multiple surgeries are scheduled. Hence, the earlier surgeries suffer more OR workload pressure than the later ones. We formally present our conjectures about the effects of orderings in Hypothesis 3.
\begin{itemize}
  \item[\textbf{Hypothesis 3:}] The mean of surgery duration decreases (increases), if a surgery takes place earlier (later) in the sequence of surgeries performed by a surgeon.
\end{itemize}
The corresponding \textit{null hypothesis} is that the mean of surgery duration does not decrease or increase with its position in the sequence of surgeries performed by a surgeon.
\subsection{Hypothesis 4: the Effects of Surgeons}
Though the surgery duration mainly depends on surgery type, we suspect that different surgeons exert different performance. For example, some surgeons are senior specialists who may need less time than junior surgeons for the same procedures. Surgery duration may also be linked to a surgeon's personal style, or preference for the use of certain surgical tools. We present the hypothesis as follows.
\begin{itemize}
  \item [\textbf{Hypothesis 4:}] The mean of surgery duration is related to the surgeon performing the procedure.
\end{itemize}
The corresponding \textit{null hypothesis} is that the mean of surgery duration has nothing to do with the surgeon performing the procedure.
\subsection{Hypothesis 5: the Effects of Interactions}
We conjecture that  surgeons are not homogeneous. Firstly, a surgeon may be good at certain types of surgeries, but less so at others, which indicates a potential interaction between surgeon and surgery type. Secondly, different surgeons may have different effects of orderings, \textit{i.e.}, an interaction between surgeons and orderings. Additionally, other interactions may also be argued, \textit{e.g.}, the interaction between orderings and surgery types, hence we will investigate the effects of interactions. We present the hypothesis formally as follows.
\begin{itemize}
  \item [\textbf{Hypothesis 5:}] The interactions between some factors, \textit{e.g.}, surgeons, orderings and surgery types, influence surgery duration.
\end{itemize}
The corresponding \textit{null hypothesis} is that there are no interactions of surgeons, orderings and surgery types that affect mean of surgery duration.

\section{Data Collection and Description} \label{sec:dataDescr}
Our data come from a large hospital in China from January 2014 through October 2016. There are 46 ORs in the hospital, which together host more than 20,000 surgeries a year. We describe the data structure from the following perspectives.

\begin{table}[h]
	\centering
	\caption{Surgery statistics I: Frequency distribution of the number of surgeries performed by each surgeon in a day}
	\label{tab:sta1}
	\begin{tabular}{crrrrrrrrr}
		\toprule[1.5pt]
		\multirow{2}{*}{Surgeon} & \multicolumn{1}{c}{Number}       & \multicolumn{1}{c}{\multirow{2}{*}{Persent}} & \multicolumn{1}{c}{} & \multicolumn{6}{c}{Number of surgeries in a day}                                                                                              \\ \cmidrule[0.7pt]{5-10}
		& \multicolumn{1}{c}{of surgeries} & \multicolumn{1}{c}{}                         & \multicolumn{1}{c}{} & \multicolumn{1}{c}{1} & \multicolumn{1}{c}{2} & \multicolumn{1}{c}{3} & \multicolumn{1}{c}{4} & \multicolumn{1}{c}{5} & \multicolumn{1}{c}{6} \\
		\midrule[1pt]                         
		A                        & 919                              & 37.49\%                                      &                      & 128                   & 308                   & 332                   & 143                   & 8                     & 0                     \\
		&                                  &                                              &                      & 13.93\%               & 33.51\%               & 36.13\%               & 15.56\%               & 0.87\%                & 0.00\%                \\
		B                        & 507                              & 20.69\%                                      &                      & 150                   & 200                   & 89                    & 52                    & 10                    & 6                     \\
		&                                  &                                              &                      & 29.59\%               & 39.45\%               & 17.55\%               & 10.26\%               & 1.97\%                & 1.18\%                \\
		C                        & 484                              & 19.75\%                                      &                      & 98                    & 265                   & 103                   & 8                     & 10                    & 0                     \\
		&                                  &                                              &                      & 20.25\%               & 54.75\%               & 21.28\%               & 1.65\%                & 2.07\%                & 0.00\%                \\
		D                        & 296                              & 12.08\%                                      &                      & 104                   & 119                   & 56                    & 15                    & 1                     & 1                     \\
		&                                  &                                              &                      & 35.14\%               & 40.20\%               & 18.92\%               & 5.07\%                & 0.34\%                & 0.34\%                \\
		E                        & 170                              & 6.94\%                                       &                      & 103                   & 56                    & 11                    & 0                     & 0                     & 0                     \\
		\multicolumn{1}{l}{}     &                                  &                                              &                      & 60.59\%               & 32.94\%               & 6.47\%                & 0.00\%                & 0.00\%                & 0.00\%                \\
		F                        & 75                               & 3.06\%                                       &                      & 21                    & 29                    & 21                    & 4                     & 0                     & 0                     \\
		\multicolumn{1}{l}{}     &                                  &                                              &                      & 28.00\%               & 38.67\%               & 28.00\%               & 5.33\%                & 0.00\%                & 0.00\%                \\
		\midrule[1pt]
		\multicolumn{1}{l}{}     & 2451                             &                                              &                      & 604                   & 977                   & 612                   & 222                   & 29                    & 7                     \\
		\multicolumn{1}{l}{}     &                                  &                                              &                      & 24.64\%               & 39.86\%               & 24.97\%               & 9.06\%                & 1.18\%                & 0.29\%               \\
		\bottomrule[1.5pt]
	\end{tabular}
\end{table}

\subsection{Surgeons}
We limit our data to one department: Thoracic Surgery department which performs the largest number of surgeries during our data collection period. We focus on 6 surgeons who performed the most surgeries. The number of surgeries performed by each surgeon is listed in the second column of Table \ref{tab:sta1}, in which the anonymous  surgeons are denoted by letters, ``A", ``B", ``C", ``D", ``E'' and ``F". Surgeon A, B and C perform the most number of surgeries, which take up more than $75\%$ of the total number.

\begin{figure}[h]
\centering
\begin{tabular}{cc}
  \includegraphics[width = 8cm]{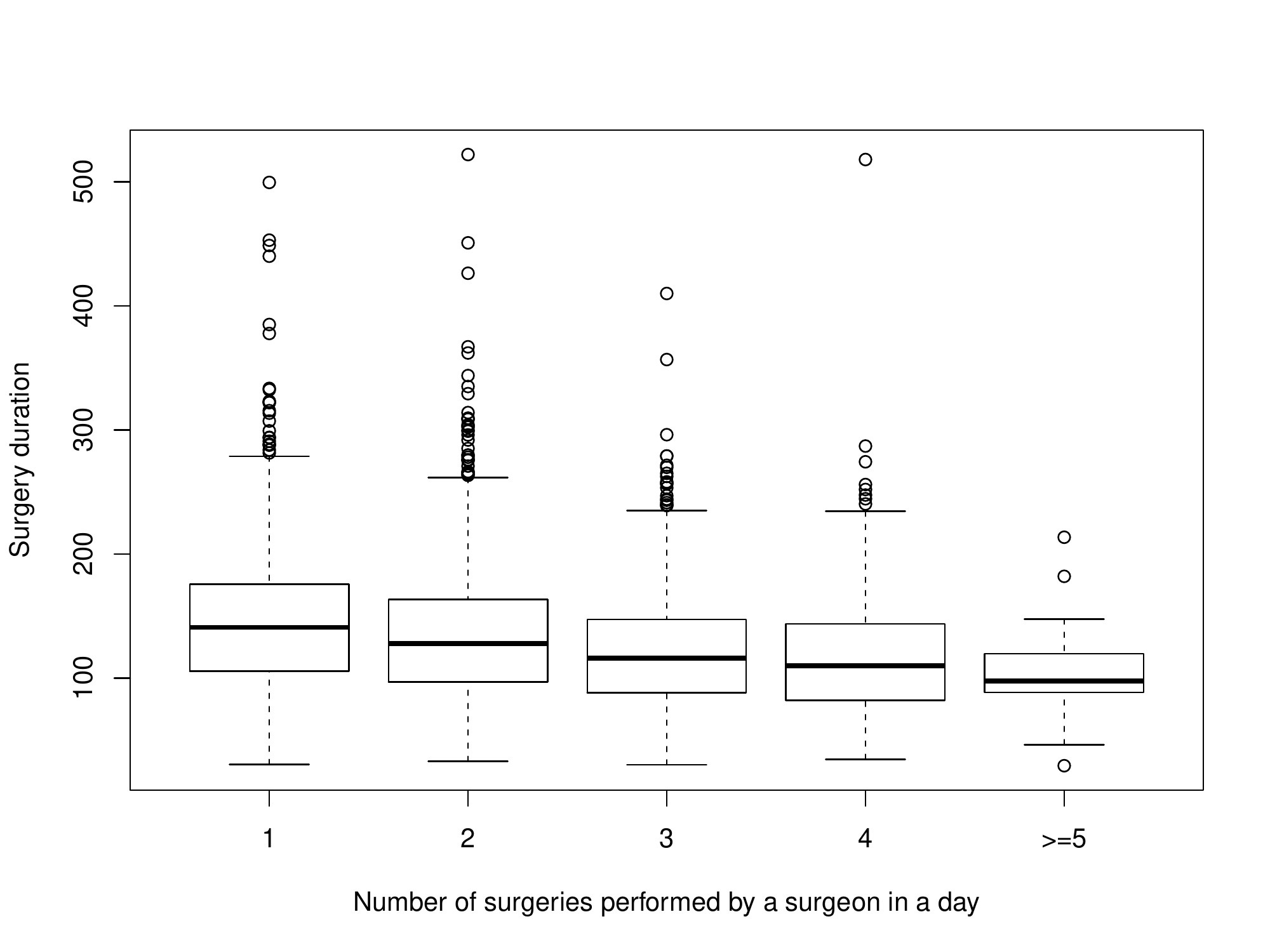} &   \includegraphics[width=8cm]{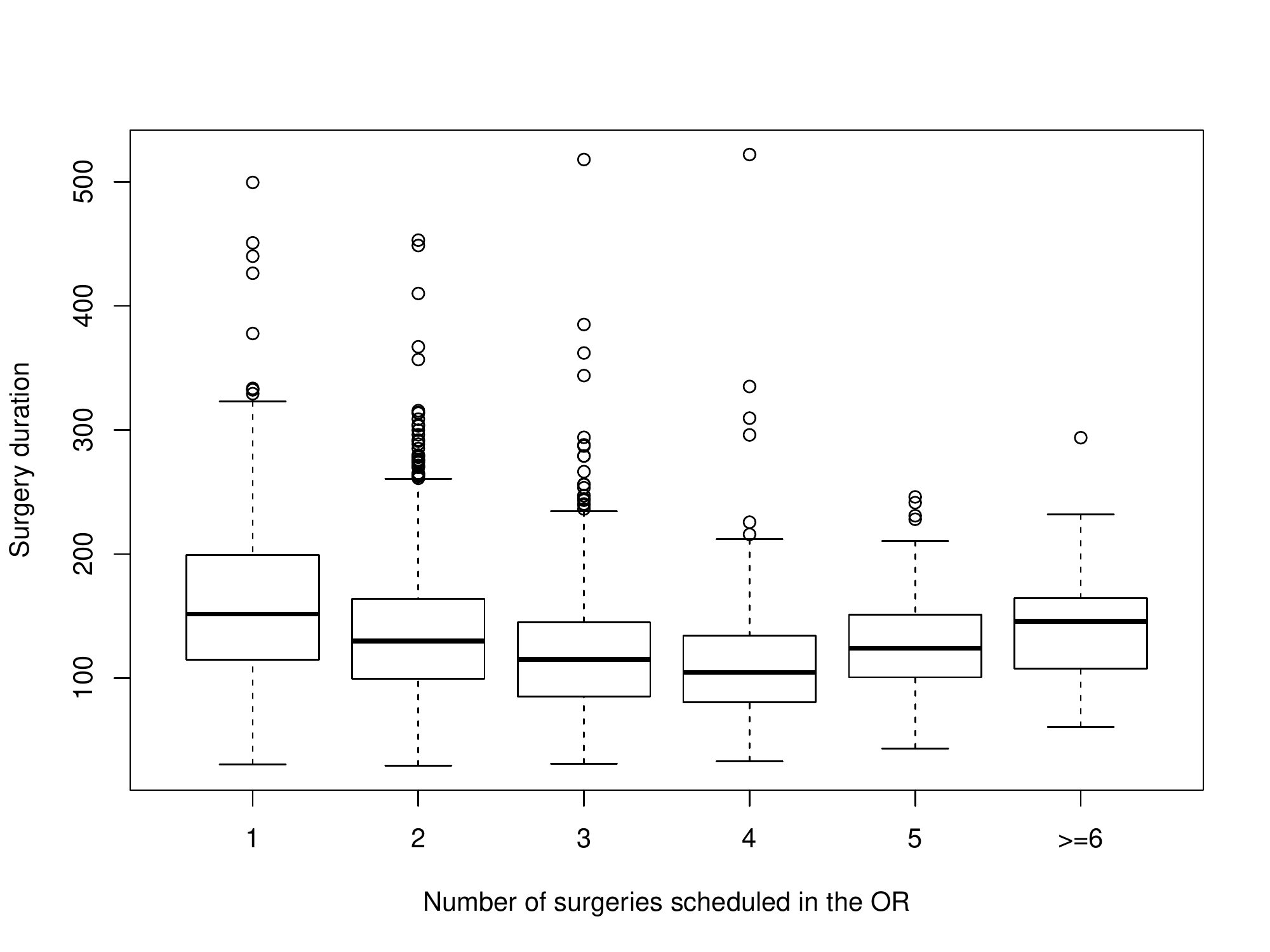} \\
 \text{\footnotesize (a)} & \text{\footnotesize (b)} \\
 \includegraphics[width=8cm]{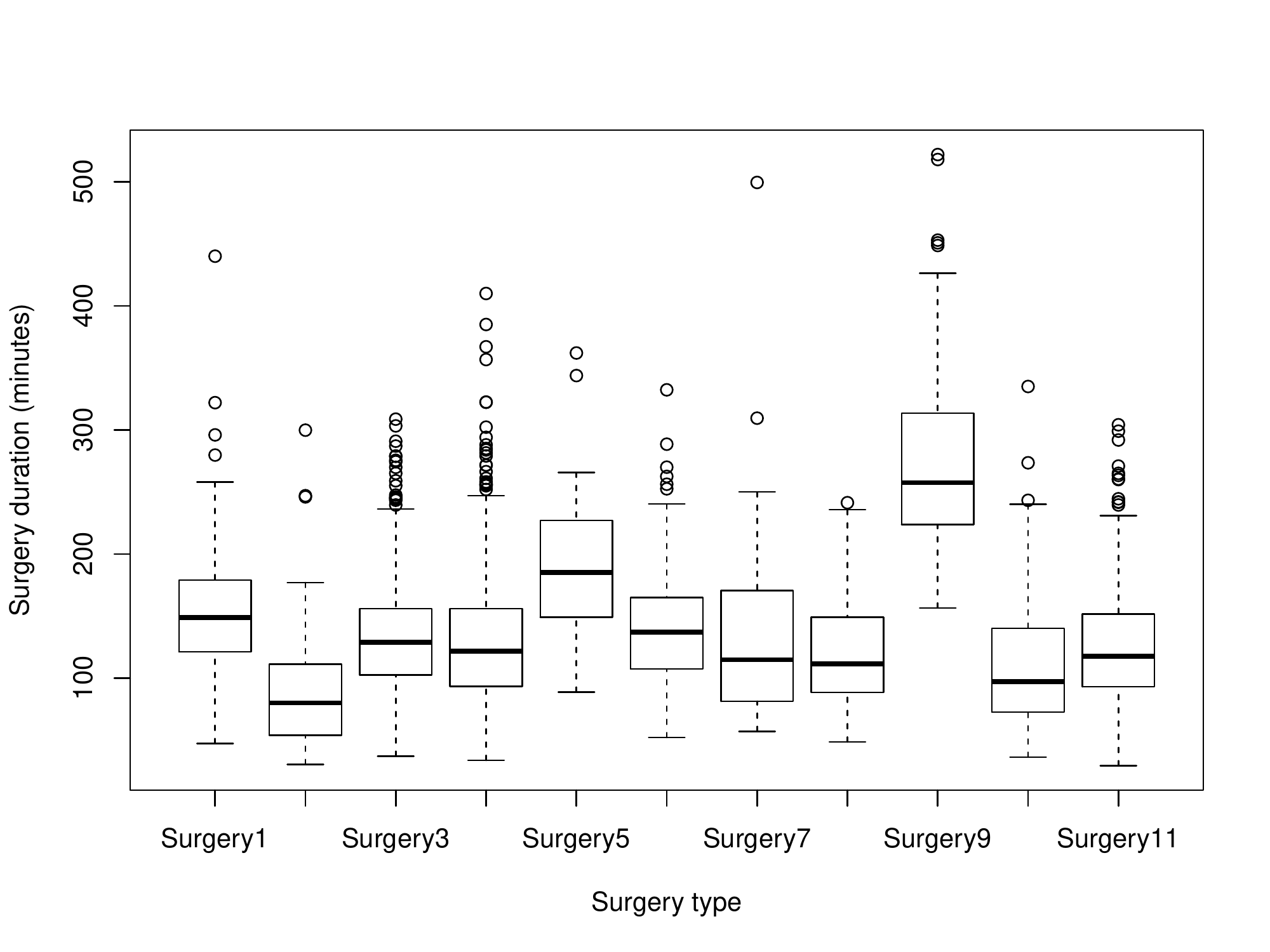} &   \includegraphics[width=8cm]{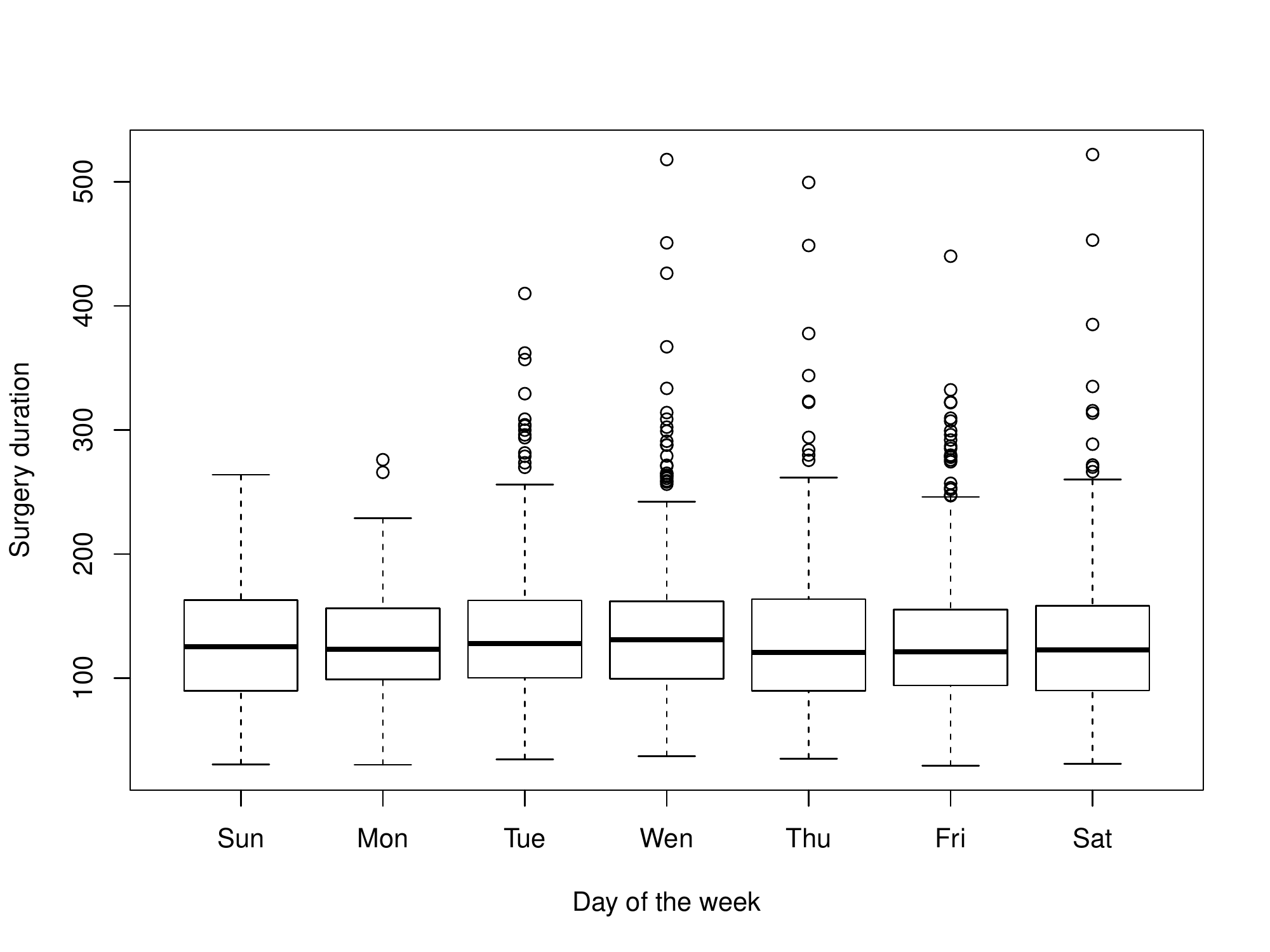} \\
 \text{\footnotesize (c) } & \text{\footnotesize (d)}
\end{tabular}
\caption{\label{fig:surgeyDur} Relationship between surgery duration and other surgery factors such as  (a) Surgery type, (b) Surgeon workload (c) Workload in OR (d) Weekday/weekend}
\end{figure}

\subsection{Number of Surgeries a Surgeon Performs in a Day}

A surgeon may perform multiple surgeries in a day. For the surgeons in consideration, the number a surgeon performs in a day varies from 1 to 6. The columns labelled 1-6  in Table \ref{tab:sta1} illustrates these statistics. The labels in the title row ($1, 2, \cdots, 6$) represent the number of surgeries that are performed by the surgeon in a day. Take the number in the row labelled ``Surgeon A'' and the column labelled ``2'', 308, for an example. It means that there are 308 surgeries each of which is performed by Surgeon A in the day with 2 surgeries finished by the surgeon. The last row lists the sum of the above numbers in each column. Note that the number in the last row might not be divisible by the number in the title row. This is because some instances are not included due to erroneous or incomplete records. In most cases surgeons perform fewer than four surgeries in a day. Also, different surgeons show different patterns. Specifically, Surgeon A performs two or three surgeries a day, while in most of the cases Surgeon B, C, D, and F tend to perform two surgeries in a day, especially for Surgeon B. Surgeon E is the only  surgeon who usually performs only one surgery in a day. Figure \ref{fig:surgeyDur}(a) demonstrates 6 box-plots of surgery durations, corresponding to the cases in which 1-6 surgeries are performed in one day, respectively. There are only a few instances in which six or more surgeries are performed in one day, hence six or more is combined as such in the sixth box-plot of Figure \ref{fig:surgeyDur}(a). It is shown that surgery duration decreases with the number of surgeries performed by a surgeon in a day. This is reasonably expected, since surgeons may accelerate if they face higher work pressure, as was also shown in \cite{Kc2009}. Also, there are many outliers if a surgeon performs two or three surgeries in a day, because there are many cases for the two scenarios. In our model, we exclude the cases with six or more surgeries a day, since there are only seven such cases, representing only 0.29\% of the data.

\subsection{Number of Surgeries Scheduled in An OR}\label{sec:numInOR}
\begin{table}[h]
\centering
\caption{Surgery statistics II: Frequency distribution of the number of surgeries scheduled in ORs performed by each surgeon }
\label{tab:numInOR}
\begin{tabular}{crrrrrrrrrr}
\toprule[1.5pt]
\multicolumn{1}{l}{\multirow{2}{*}{Surgeon}} & \multicolumn{1}{c}{Number}       & \multicolumn{1}{c}{\multirow{2}{*}{Percent}} & \multicolumn{1}{c}{} & \multicolumn{7}{c}{Number of surgeries in an OR in a day}                                                                                                             \\ \cmidrule[0.7pt]{5-11}
\multicolumn{1}{l}{}                         & \multicolumn{1}{c}{of surgeries} & \multicolumn{1}{c}{}                         & \multicolumn{1}{c}{} & \multicolumn{1}{c}{1} & \multicolumn{1}{c}{2} & \multicolumn{1}{c}{3} & \multicolumn{1}{c}{4} & \multicolumn{1}{c}{5} & \multicolumn{1}{c}{6} & \multicolumn{1}{c}{7} \\
\midrule[1pt]
A                                            & 919                              & 37.49\%                                      &                      & 76                    & 473                   & 284                   & 55                    & 23                    & 8                     & 0                     \\
                                             &                                  &                                              &                      & 8.27\%                & 51.47\%               & 30.90\%               & 5.98\%                & 2.50\%                & 0.87\%                & 0.00\%                \\
B                                            & 507                              & 20.69\%                                      &                      & 36                    & 236                   & 140                   & 73                    & 14                    & 5                     & 3                     \\
                                             &                                  &                                              &                      & 7.10\%                & 46.55\%               & 27.61\%               & 14.40\%               & 2.76\%                & 0.99\%                & 0.59\%                \\
C                                            & 484                              & 19.75\%                                      &                      & 43                    & 290                   & 103                   & 31                    & 10                    & 5                     & 2                     \\
                                             &                                  &                                              &                      & 8.88\%                & 59.92\%               & 21.28\%               & 6.40\%                & 2.07\%                & 1.03\%                & 0.41\%                \\
D                                            & 296                              & 12.08\%                                      &                      & 59                    & 174                   & 47                    & 11                    & 1                     & 4                     & 0                     \\
                                             &                                  &                                              &                      & 19.93\%               & 58.78\%               & 15.88\%               & 3.72\%                & 0.34\%                & 1.35\%                & 0.00\%                \\
E                                            & 170                              & 6.94\%                                       &                      & 31                    & 86                    & 31                    & 17                    & 3                     & 1                     & 1                     \\
                                             &                                  &                                              &                      & 18.24\%               & 50.59\%               & 18.24\%               & 10.00\%               & 1.76\%                & 0.59\%                & 0.59\%                \\
F                                            & 75                               & 3.06\%                                       &                      & 11                    & 37                    & 27                    & 0                     & 0                     & 0                     & 0                     \\
                                             &                                  &                                              &                      & 14.67\%               & 49.33\%               & 36.00\%               & 0.00\%                & 0.00\%                & 0.00\%                & 0.00\%                \\
\midrule[1pt]                                             
                                             & 2451                             &                                              &                      & 256                   & 1296                  & 632                   & 187                   & 51                    & 23                    & 6                     \\
                                             &                                  &                                              &                      & 10.44\%               & 52.88\%               & 25.79\%               & 7.63\%                & 2.08\%                & 0.94\%                & 0.24\%  \\
\bottomrule[1.5pt]                                                          
\end{tabular}
\end{table}

Similar to the illustration in Table \ref{tab:sta1}, Table \ref{tab:numInOR} illustrates the number of surgeries scheduled in the OR in a day. It is shown that in most cases there are no more than three surgeries in an OR, and surgeons have similar patterns that most of the surgeries are done in an OR with two surgeries in a day. Figure \ref{fig:surgeyDur}(b) is the box-plot of the surgery duration under a different number of surgeries in an OR. As in Figure \ref{fig:surgeyDur}(a), we combine the surgeries performed in an OR of seven or more in a day. There is a tipping point. Specifically, in the cases that there are four or fewer surgeries in an OR, the surgery duration decreases with the number of surgeries, while the duration increases if the number of surgeries in an OR is five or more. We will test the existence of a tipping point later.

\subsection{Surgery Types}
\begin{table}[h]
\centering \small
\caption{Summary statistics of surgery duration for surgery types}
\label{tab:DurSta}
\begin{tabular}{clrrrrrr}
\toprule[1.5pt]
Index & Surgery Name                                 & Frequency & Percent  & Median & Min    & Mean   & Max    \\\midrule[1pt]
1 & Surgery for lung cancer & 99     & 4.04\%   & 148.85 & 47.17  & 154.05 & 440.05 \\
2 & Thoracoscopic pulmonary bullous resection & 61     & 2.49\%   & 80.00  & 30.33  & 92.05  & 299.83 \\
3 & Thoracoscopic partial pulmonary lobectomy & 443    & 18.07\%  & 128.62 & 37.00  & 133.14 & 308.70 \\
4 & Thoracoscopic interior pulmonary lobectomy & 930    & 37.94\%  & 121.60 & 33.67  & 129.12 & 410.00 \\
5 & Total pneumonectomy   & 27     & 1.10\%   & 185.00 & 88.67  & 194.05 & 362.03 \\
6 & Partial pulmonary lobectomy  & 280    & 11.42\%  & 136.93 & 52.00  & 140.54 & 332.33 \\
7 & Thoracoscopic exploration & 32     & 1.31\%   & 114.75 & 57.00  & 145.58 & 499.52 \\
8 & Pulmonary wedge resection     & 58     & 2.37\%   & 111.30 & 48.50  & 122.31 & 241.37 \\
9 & Oesophageal cancer      & 50     & 2.04\%   & 257.46 & 156.52 & 278.82 & 522.00 \\
10 & Mediastinal tumour resection  & 142    & 5.79\%   & 97.09  & 36.17  & 111.02 & 335.03 \\
11 & Pulmonary tumour resection & 329    & 13.42\%  & 117.72 & 29.33  & 125.54 & 304.17 \\\midrule[1pt]
 &       & 2451   & 100.00\% & 125.00 & 29.33  & 133.53 & 522.00\\
 \bottomrule[1.5pt]
\end{tabular}
\end{table}

The surgeons in the department are specialized in thoracic surgeries, which can be categorized into 11 types. Table \ref{tab:DurSta} presents surgery types and the corresponding statistics. The fourth type of surgery has the highest occurrence and is the main surgical procedure that the department performs.  Figure \ref{fig:surgeyDur}(c) shows the box plot of surgery duration for each surgery type. Surgery 9 (oesophageal cancer) has the longest duration. The duration of another cancer, lung cancer (Surgery 1), is also longer than most surgeries. For the main surgery in the department (pulmonary lobectomy), total pneumonectomy (Surgery 5) is reasonably more time-consuming than partial/interior pulmonary lobectomy (Surgery 3 and 4). The mean of Surgery 2 duration is the smallest. The duration of Surgery 3 and 4 have similar means and medians. They also have the highest number of outliers, which is may be due to their high occurrence.

\subsection{Number of Surgeries in the day of the week}\label{sec:weekday}
\begin{table}[h]
\centering
\caption{Number surgeries performed in the day of the week}
\label{tab:weekday}
\begin{tabular}{crr}
\toprule[1.5pt]
Weekday & Number & Percent  \\
\midrule[1pt]
Sun     & 30     & 1.22\%   \\
Mon     & 44     & 1.80\%   \\
Tue     & 509    & 20.77\%  \\
Wen     & 491    & 20.03\%  \\
Thu     & 435    & 17.75\%  \\
Fri     & 563    & 22.97\%  \\
Sat     & 379    & 15.46\%  \\
\midrule[1pt]
        & 2451   & 100.00\%\\
\bottomrule[1.5pt]       
\end{tabular}
\end{table}

 We illustrate the number of surgeries in each day of the week in Table \ref{tab:weekday}, which shows that there are only a small portion of surgeries scheduled on Sunday and Monday. Hence, to some extent, Sunday and Monday represent the ``weekends'' for the surgeons. Figure \ref{fig:surgeyDur}(d) shows the box-plot of surgery duration for each day of the week. The mean of surgery duration does not seem significantly different for different days of the week.

\section{Statistical Modelling and Analysis}\label{sec:ana}
In this section, we propose statistical models to test the hypotheses developed in Section \ref{sec:hypo}.
\subsection{Variables}
The variables in consideration are listed in Table \ref{tab:variables}. The terms, ``variable'', ``predictor'', and ``factor'', are used interchangeably throughout the paper.
We divide the variables into two categories: clinical variables and non-clinical variables. Clinical variable is $SURGRYTYPE$, while the non-clinical variables include $SURGEON$, $ANESTHETIST$, $DAY$, $W\_SURGEON$ (the number of surgeries the surgeon performs in a day), $ORDER$ (the position of a surgery in the sequence of the surgeries performed by the surgeon in a day) and $W\_OR$ (the number of surgeries scheduled in the OR). $ORDER$ is a categorical variable, taking the value of $ONLYONE$ if a surgery is the only surgery a surgeon performed in a day, ``$2\sim1$'' if the surgery is the first surgery that is done by a surgeon who performs two surgeries on that day, ``$2\sim 2$'', ``$3\sim 1$'', and so forth.

\begin{table}[h]
	\centering
	\caption{\label{tab:variables} Description of variables used in the models}
	\label{tab:varIntro}
	\begin{tabular}{r|p{12cm}}
		\toprule[1.5pt]
		\textbf{Variable} & \textbf{Description}\\
		\midrule[1pt]
		$SURGEON$ & the surgeon performing the procedure\\
		$ANESTHETIST$ & the surgery anesthetist \\
		$SURGRYTYPE$ & the surgery type, one of 11 types shown in Table \ref{tab:DurSta} \\
		$DAY$ & the day of the week when the surgery is performed\\
		$W\_SURGEON$ & the number of surgeries the surgeon performs in a day\\
		$ORDER$ & the position of a surgery in the sequence of surgeries performed by the surgeon in a day \\
		$W\_OR$ & the number of surgeries scheduled in the OR where the surgery is performed\\
		\bottomrule[1.5pt]       
	\end{tabular}
\end{table}
\subsection{Simple Models}
We first use simple models (the models without interactions) to test Hypothesis 1-4. In this paper, we regard the effects of the predictors listed in Table \ref{tab:variables} as the main effects. In Figure 1 we observe that the the response variable is very skewed. This is against the assumptions that we make when  we fit a linear model and has negative impact in the estimation of the model. To avoid this problem, we take the log of surgery duration and use it as the response. Since the variables $ORDER$ and $W\_SURGEON$ are collinear, the following models (Model I and II) with the two variables are developed respectively.
\begin{align*}
	& \begin{array}{rl}
		\textbf{Model I: }log(DURATION_i) =& \beta_0 + \beta_1 DAY_i+  \beta_2 W\_SURGEON_i + \beta_{4} W\_OR_i + \beta_5 SURGEON_i\\
		   &+ \beta_6  SURGERYTYPE_i + \beta_7 ANESTHETIST_i 
	\end{array}\\
	& \begin{array}{rl}
		\textbf{Model II: }log(DURATION_i) =& \beta_0 + \beta_1 DAY_i+  \beta_3 ORDER_i + \beta_{4} W\_OR_i + \beta_5 SURGEON_i\\
		 &+ \beta_6  SURGERYTYPE_i + \beta_7 ANESTHETIST_i
	\end{array}
\end{align*}
Inspired by Figure \ref{fig:surgeyDur}(c), we would like to test different effects of $W\_OR$ below and above the tipping point (denoted as $TP$). Hence, we use the following piecewise linear function to substitute $\beta_{4} W\_OR_i$ in Model I and II. 
\begin{equation*}
	\beta_{41} W\_OR1_i + \beta_{42} W\_OR2_i  = \beta_{41} W\_OR_i + \beta_{42} \max\{W\_OR_i - TP,0\}
\end{equation*}
 which indicates that $\beta_{41}$ estimates the slope below the tipping point, and $\beta_{41}+\beta_{42}$ estimates the slope above the tipping point. Thereby we have the following two models.
\begin{align*}
	& \begin{array}{rl}
		\textbf{Model III: }log(DURATION_i) =& \beta_0 + \beta_1 DAY_i+  \beta_2 W\_SURGEON_i+\beta_{41}\min\{ W\_OR_i,TP\}\\
		  &+ \beta_{42} \max\{W\_OR_i - TP,0\}+ \beta_5 SURGEON_i\\ 		 
		  &+ \beta_6  SURGERYTYPE_i + \beta_7 ANESTHETIST_i
	\end{array}\\
	& \begin{array}{rl}
		\textbf{Model IV: }log(DURATION_i) =& \beta_0 + \beta_1 DAY_i+ \beta_3 ORDER_i+\beta_{41}\min\{ W\_OR_i,TP\}\\
		  &+ \beta_{42} \max\{W\_OR_i - TP,0\}+ \beta_5 SURGEON_i \\ 
		  &+ \beta_6  SURGERYTYPE_i + \beta_7 ANESTHETIST_i
	\end{array}
\end{align*} 

The regression results are illustrated in Table \ref{tab:simM}. The terms relevant to the factor $ANESTHETIST$ are not included because it has 50 categories and it will add too many lines to the table. The factor $SURGERYTYPE$ is also excluded since by using the simple models we would like to test Hypotheses 1-4, which do not refer to surgery types. 

Hypothesis 1 is supported by all models. To show the effect of the day of the week, we set Saturday as the baseline. In all models, the effects of Tuesday and Friday are positively significant, indicating that duration of surgeries on these two days in a week is longer than those on Saturday. If we compare Table \ref{tab:weekday} and Table \ref{tab:simM}, we see that the two busiest days, which are Tuesdays and Fridays, have slightly longer average operation time. They are followed by Wednesday and then the other week days. Besides, as mentioned in Section \ref{sec:weekday}, Sunday and Monday represent the weekends for the surgeons. Hence, longer duration in Tuesday might reveal a weekday effect, \textit{i.e.}, surgeons appear to work at a lower pace when they go back to work after the weekend. Numerically, the duration of surgeries on Tuesday or Friday is around 8 minutes longer than those on Saturday.

\begin{table}[h]
\centering
\caption{Regression results for models using only the main effect and without interaction terms (simple models). Anesthetist and surgery type were excluded because of space.}
\label{tab:simM}
\begin{tabular}{lllll}
\toprule[1.5pt]
                             & \multicolumn{4}{c}{Coefficient}                                                                               \\\cmidrule[0.7pt]{2-5}
Variable                    & Model I                   & Model II                  & Model III                 & Model IV                  \\ \midrule[1pt]
Sunday & -0.0294 & -0.0313 & -0.0412 & -0.0445\\
Monday & 0.0557 & 0.0661 & 0.0652 & -0.0766\\
Tuesday & 0.0600* & 0.0613* & 0.0561* & 0.0572* \\
Wednesday & 0.0418 & 0.0442 & 0.0328 & 0.0348\\
Thursday & 0.0072 & 0.0084 & 0.0074 & 0.0087\\
Friday & 0.0552* & 0.0544* & 0.0543* & 0.0533*\\
Order 2$\sim$1               &                           & -0.081***                 &                           & -0.0748**                 \\
Order 2$\sim$2               &                           & -0.0756**                 &                           & -0.0669**                 \\
Order 3$\sim$1               &                           & -0.1835***                &                           & -0.1693***                \\
Order 3$\sim$2               &                           & -0.1558***                &                           & -0.1452***                \\
Order 3$\sim$3               &                           & -0.1722***                &                           & -0.1631***                \\
Order 4$\sim$1               &                           & -0.2343***                &                           & -0.206***                 \\
Order 4$\sim$2               &                           & -0.1688**                 &                           & -0.1413*                  \\
Order 4$\sim$3               &                           & -0.1528**                 &                           & -0.1289*                  \\
Order 4$\sim$4               &                           & -0.1752**                 &                           & -0.1483**                 \\
Order 5$\sim$1               &                           & -0.3204*                  &                           & -0.2874.                  \\
Order 5$\sim$2               &                           & -0.08                     &                           & -0.051                    \\
Order 5$\sim$3               &                           & -0.1702                   &                           & -0.1318                   \\
Order 5$\sim$4               &                           & -0.4697**                 &                           & -0.4426*                  \\
Order 5$\sim$5               &                           & -0.1469                   &                           & -0.1055                   \\
W\_SURGEON                   & -0.0686***                &                           & -0.0613***                &                           \\
W\_OR1                       &                           &                           & -0.0662***                & -0.0667***                \\
W\_OR2                       &                           &                           & 0.1033**                  & 0.1093**                  \\
W\_OR                        & -0.0377***                & -0.0371***                &                           &                           \\
Surgeon B                    & -0.0978***                & -0.1016***                & -0.0936***                & -0.0977***                \\
Surgeon C                    & 0.0569*                   & 0.0577*                   & 0.0579*                   & 0.0586*                   \\
Surgeon D                    & -0.0037                   & -0.0054                   & -0.006                    & -0.0081                   \\
Surgeon E                    & -0.039                    & -0.0433                   & -0.0345                   & -0.0391                   \\
Surgeon F                    & -0.2761***                & -0.2746***                & -0.2722***                & -0.2704***                \\\midrule[0.5pt]
Log likelihood               &\multicolumn{1}{c}{-1079}                     & \multicolumn{1}{c}{-1073} & \multicolumn{1}{c}{-1070}   & \multicolumn{1}{c}{-1064}   \\ 
Adjusted $R^2$ & \multicolumn{1}{c}{0.198} & \multicolumn{1}{c}{0.197} & \multicolumn{1}{c}{0.203} & \multicolumn{1}{c}{0.203}\\ 
\bottomrule[1.5pt]
\end{tabular}
\end{table}

 Hypothesis 2(1) is supported by Model I and III, \textit{i.e.}, surgery duration decreases with the number of surgeries a surgeon performs in a day. On average, surgery duration will decrease by 10 minutes if a surgeon performs one more surgery. It seems that Hypothesis 2(2) is supported by Model I and II, \textit{i.e.}, surgery duration decreases with the number of surgeries scheduled in an OR. However, Model III and IV with tipping point demonstrate a better fit, with larger log likelihood and adjusted $R^2$ values. Hence, we can conclude that a tipping point of the workload in OR exists. That is, surgery duration decreases with the number of surgeries allocated to an OR if there are no more than 4 surgeries in the OR, while the duration increases with the number if there are more than 4 surgeries allocated to the OR. As we mentioned in Section \ref{sec:hypo-2}, high workload in an OR results in decreased surgery duration. However, the regression result demonstrates that surgery duration will increase if the number of surgeries scheduled in an OR in a day is beyond a certain value (say 4 for our data set). One possible reason is that if too many surgeries are allocated in an OR, the OR becomes disordered, which results in longer surgery duration. Numerically, when there are no more than 4 surgeries in an OR, surgery duration decreases around 8 minutes if one more surgery is allocated into the OR; and when there are more than 4 surgeries in an OR, surgery duration increases around 12 minutes if one more surgery is added to the OR.
 
\begin{figure}[h]
\centering
\begin{centering}
\subfloat[]{\begin{centering}
\includegraphics[width=8cm]{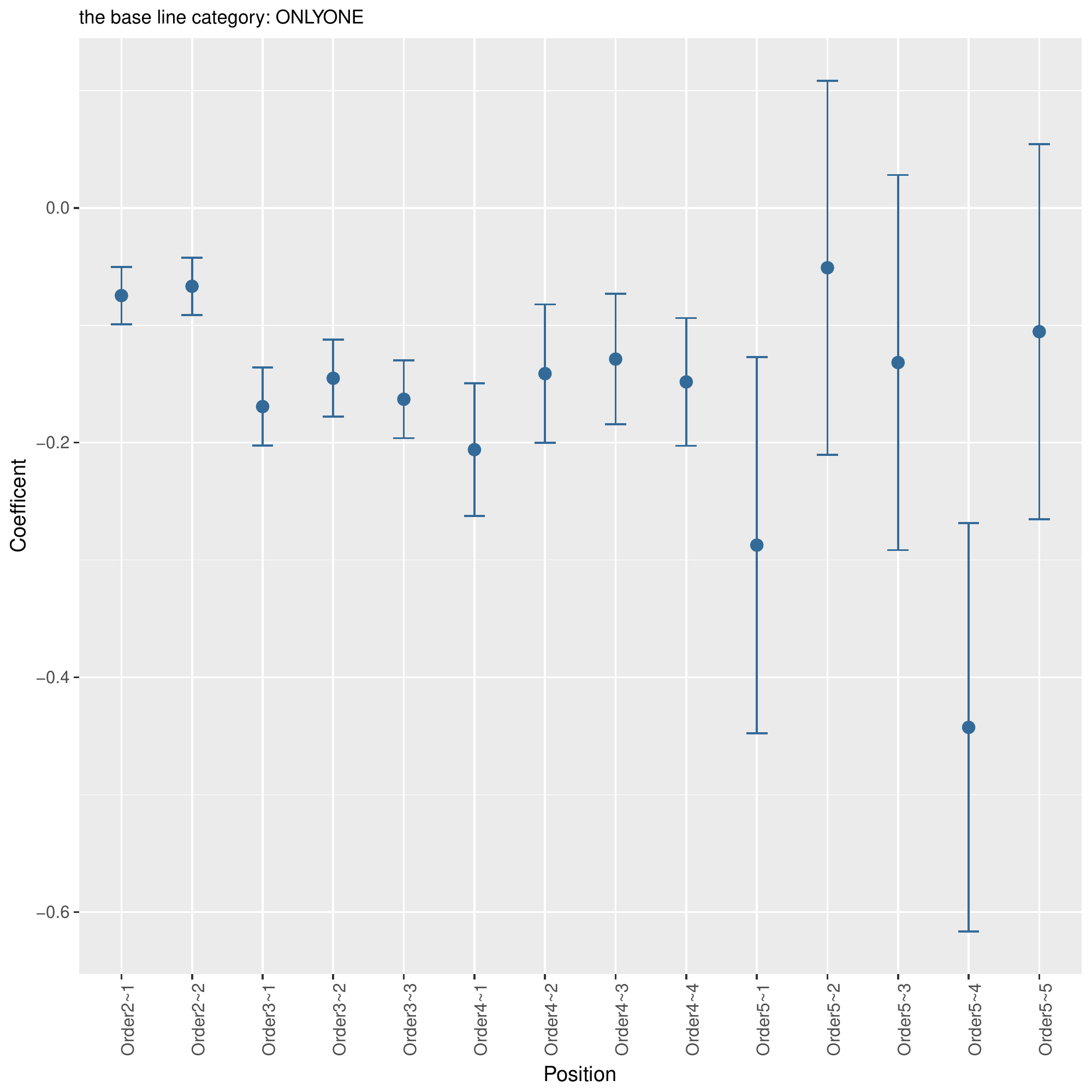}
\par\end{centering}
\label{fig:coeffOrder}}
\subfloat[]{\begin{centering}
\includegraphics[width=8cm]{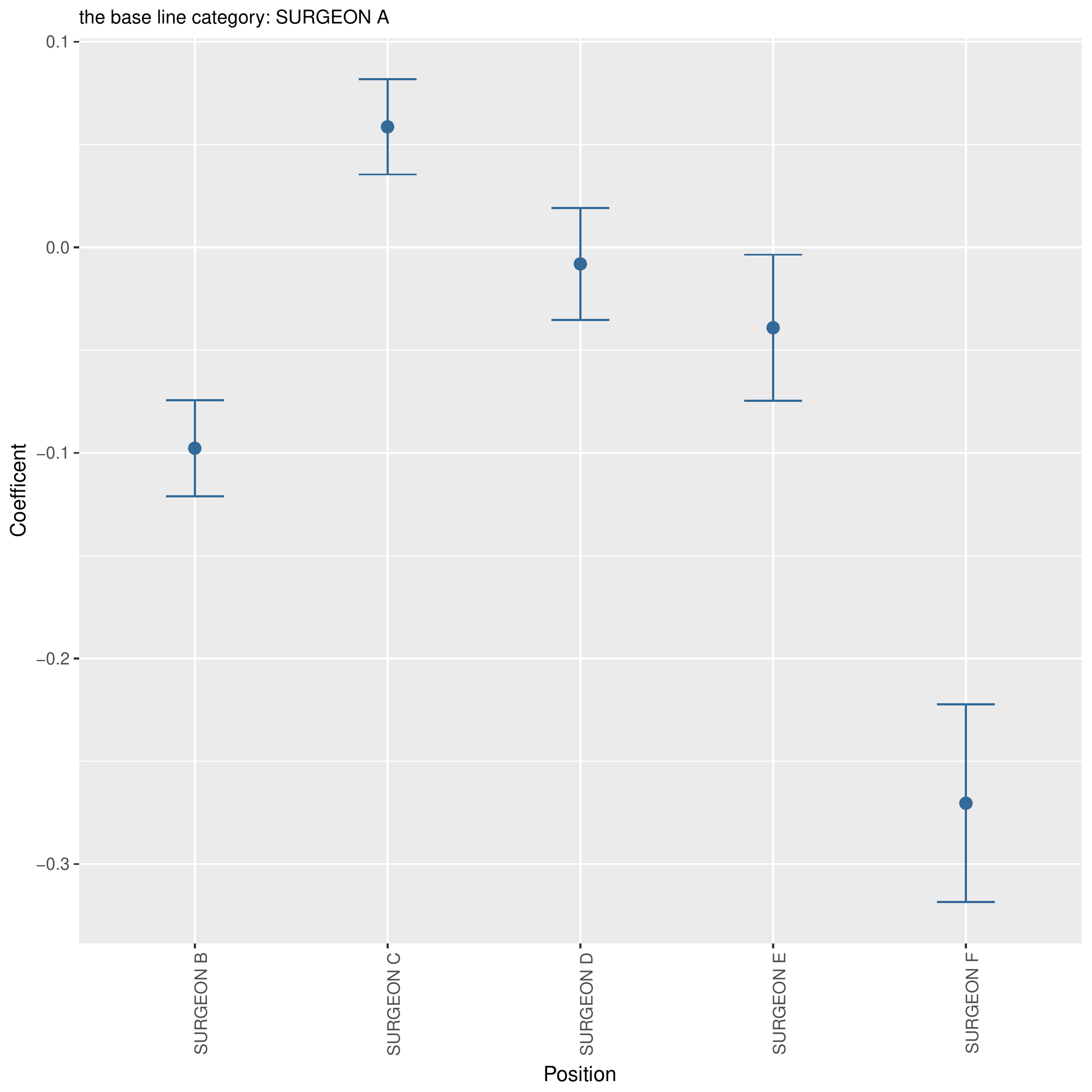}
\par\end{centering}
\label{fig:coeffSurgeon}
 }
\par\end{centering}
\caption{\label{fig:coeff}Estimation of (a) Surgeon effects and (b) Ordering position effects with error bars representing their standard errors}	
\end{figure}

 As for Hypothesis 3, it is partially supported by Model II and IV. Though some items are not significant when a surgeon performs five surgeries in a day, they only take up 1.18\% of all cases (see Table \ref{tab:sta1}). Note that the base line is $ONLYONE$. All coefficients of orderings are negative, which also support Hypothesis 2(1). The coefficients of different orderings do not always monotonically increase, but are very close for a certain number of surgeries a surgeon performs in a day (see the coefficients of Order $3\sim \cdot$ and Order $4\sim \cdot$). We can only conclude that the duration of the first surgery decreases more than later surgeries when a surgeon has two, three or four surgeries in a day. Figure \ref{fig:coeffOrder} shows a bar plot of coefficient estimations of surgery positions. It shows that the standard errors increase with the number of surgeries performed by a surgeon in a day, especially when a surgeon performs five surgeries. For the extend to which the duration decreases, duration of surgeries in position (4, 1) suffers the biggest drop around 28 minutes (for the cases where there are no more than 4 surgeries per surgeon in a day).

Hypothesis 4 is supported by all models, \textit{i.e.}, the mean of surgery duration is related to the surgeon performing the procedure. We set Surgeon A as the baseline for the categorical variable $SURGEON$. It indicates that the effects of Surgeon B, C and F are significant. The duration of surgeries performed by Surgeon B and F (especially Surgery F) is less than that performed by Surgeon A, while the duration of Surgeon C's surgeries is longer than that of Surgeon A. Figure \ref{fig:coeffSurgeon} is the bar plot of coefficient estimations of surgeons, indicating the differences among the surgeons. Surgeon C appears to be slowest, with average surgery duration approximately 13 minutes more than that of Surgeon A. Surgeon F appears to be fastest, whose surgery duration is around 34 minutes less than that of Surgeon A, but has the largest standard error.

To summarize, in this subsection we tested Hypothesis 1-4. In other words, we investigated effects of the non-clinical factors on surgery duration, including day of the week, surgeon workload, workload in an OR, the position of the surgery, and surgeon. Specifically, Hypothesis 1, 2(1), 3, and 4 are supported by the regression results. For Hypothesis 2(2), we find that surgery duration does not monotonically decrease with the workload in the OR. Instead, there exists a tipping point. Surgery duration will increase if the workload in the OR is beyond the tipping point.
 
\subsection{Models with Interactions}
\begin{figure}[h]
	\centering
	\includegraphics[width = 10cm]{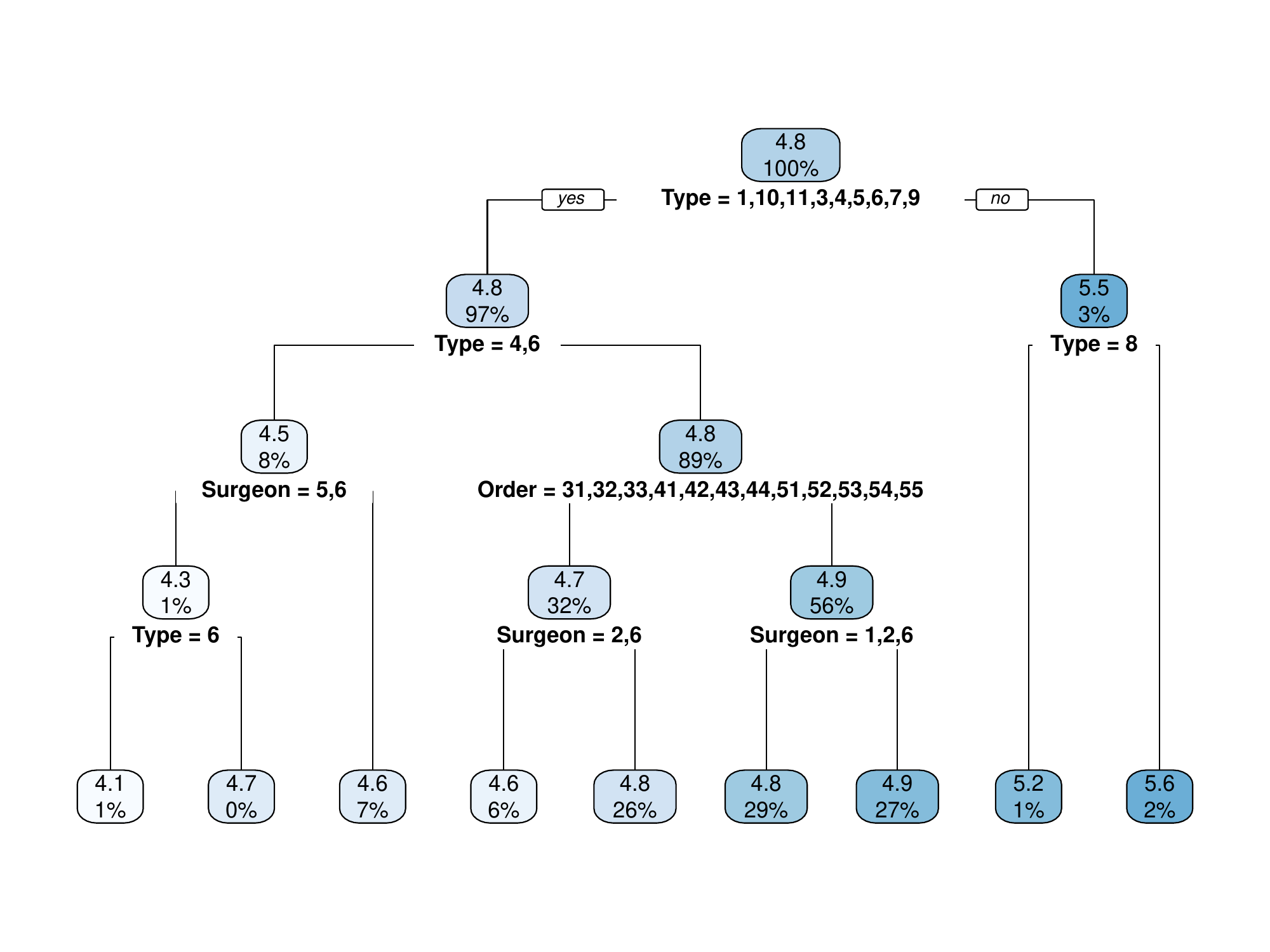}
	\caption{\label{fig:tree}Regression tree of Surgery duration by the model predictors. The objective of this tree is to uncover potential interactions terms for the model}
\end{figure}
In this subsection, we would like to test Hypothesis 5, \textit{i.e.}, the effects of some interactions on surgery duration. The question is: which interactions should be taken into account? We use the \verb|rpart()| function in R to build a regression tree of surgery duration by the model predictors. It is illustrated in Figure \ref{fig:tree}, which provides some evidence that there may be interactions among $SURGERYTYPE$, $SURGEON$ and $ORDER$. Hence, there are three kinds of interactions we need to consider.

We formulate the models with interactions based on Model IV as follows, since it is the best among the simple models. 
\begin{align*}
	& \begin{array}{rl}
		\textbf{Model V: }log(DURATION_i) = \beta_0 + \beta_1 DAY_i+ \beta_3 ORDER_i+\beta_{41}\min\{ W\_OR_i,TP\}&\\
		  + \beta_{42} \max\{W\_OR_i - TP,0\}+ \beta_5 SURGEON_i + \beta_6  SURGERYTYPE_i  &\\
		  + \gamma_1 SURGEON_i: SURGERYTYPE_i + \beta_7 ANESTHETIST_i&
	\end{array} \\
		& \begin{array}{rl}
		\textbf{Model VI: }log(DURATION_i) = \beta_0 + \beta_1 DAY_i+ \beta_3 ORDER_i+\beta_{41}\min\{ W\_OR_i,TP\}&\\
		  + \beta_{42} \max\{W\_OR_i - TP,0\}+ \beta_5 SURGEON_i + \beta_6  SURGERYTYPE_i &\\
		  + \gamma_2 SURGEON_i:ORDER_i + \beta_7 ANESTHETIST_i&
	\end{array} \\
		& \begin{array}{rl}
		\textbf{Model VII: }log(DURATION_i) = \beta_0 + \beta_1 DAY_i+ \beta_3 ORDER_i+\beta_{41}\min\{ W\_OR_i,TP\}&\\
		  + \beta_{42} \max\{W\_OR_i - TP,0\}+ \beta_5 SURGEON_i + \beta_6  SURGERYTYPE_i  &\\
		  + \gamma_3 SURGERYTYPE_i:ORDER_i + \beta_7 ANESTHETIST_i&
	\end{array} 
\end{align*} 
Next, we need to examine whether the new models are superior to the previous ones, \textit{i.e.}, whether the effects of interactions are significant. Hence, we run \verb|anova()| to perform analysis of variance for the results of Model IV and those of the new models, respectively. The results listed in Table \ref{tab:anova} indicate that the three kinds of interactions are all significant.

\begin{table}[h]
\centering
\caption{Results of analysis of variance for Model IV, V, VI, and VII}
\label{tab:anova}
\begin{tabular}{lcccccc}
\toprule[1.5pt]
& Res.Df    & RSS  & Df     & Sum of Sq & F      & Pr($\textgreater$F)        \\ 
\midrule[1pt]
Model IV  & 2351 & 341.76 &           &        &                   &         \\
Model V   & 2304 & 329.02 & 47        & 12.736 & 1.8976            & 0.0002* \\
Model VI  & 2261 & 325.85 & 90        & 15.908 & 1.2264            & 0.0761. \\
Model VII & 2299 & 330.92 & 52        & 10.841 & 1.4485            & 0.0204*\\
\bottomrule[1pt]
\end{tabular}
\end{table}

\begin{table}[h]
\centering
\caption{The regression results relevant to $SURGEON$, $SURGERYTYPE$ and their interaction obtained by using Lasso for Model V}
\label{tab:lassoModelV}
\begin{tabular}{lrrrrl}
\toprule[1.5pt]
\multicolumn{1}{c}{Variable}      & \multicolumn{1}{c}{Coefficient} & \multicolumn{1}{c}{Std. Error} & \multicolumn{1}{c}{t value} & \multicolumn{1}{c}{Pr($>|$t$|$)} & \multicolumn{1}{c}{} \\
\midrule[1pt]
Surgeon B                          & -0.1153                         & 0.0231                         & -4.986                      & 6.62e-07                                & ***                  \\
Surgeon C                          & 0.0454                          & 0.0227                         & 2.003                       & 0.0453                                & *                    \\
Surgeon F                          & -0.2138                         & 0.0501                         & -4.271                      & 2.03e-05                                & ***                  \\
SurgeryType 1                      & 0.1749                          & 0.0402                         & 4.349                       & 1.43e-05                                & ***                  \\
SurgeryType 2                      & -0.3559                         & 0.0786                         & -4.526                      & 6.31e-06                                & ***                  \\
SurgeryType 5                      & 0.4288                          & 0.0784                         & 5.468                       & 5.02e-08                                & ***                  \\
SurgeryType 6                      & 0.1027                          & 0.0257                         & 4.004                       & 6.43e-05                                & ***                  \\
SurgeryType 8                      & -0.0841                         & 0.0508                         & -1.655                      & 0.0980                                & .                    \\
SurgeryType 9                      & 0.8218                          & 0.0703                         & 11.696                      & \textless2.00e-16                          & ***                  \\
SurgeryType 10                     & -0.1654                         & 0.0337                         & -4.914                      & 9.51e-07                                & ***                  \\
SurgeryType 11                     & -0.1161                         & 0.0308                         & -3.769                      & 0.0002                                & ***                  \\
Surgeon B : SurgeryType 2           & 0.2508                          & 0.1221                         & 2.054                       & 0.0401                                & *                    \\
Surgeon B : SurgeryType 3           & 0.0895                          & 0.0567                         & 1.577                       & 0.1149                                &                      \\
Surgeon B : SurgeryType 7           & -0.4774                         & 0.1709                         & -2.793                      & 0.0053                                & **                   \\
Surgeon B : SurgeryType 11          & 0.1594                          & 0.0685                         & 2.329                       & 0.0199                                & *                    \\
Surgeon C : SurgeryType 11          & 0.3472                          & 0.2290                         & 1.517                       & 0.1295                                &                      \\
Surgeon C : SurgeryType 9           & 0.1566                          & 0.0608                         & 2.575                       & 0.0101                                & *                    \\
Surgeon D : SurgeryType 5           & -0.4320                         & 0.2309                         & -1.871                      & 0.0615                                & .                    \\
Surgeon D : SurgeryType 9           & -0.2881                         & 0.1172                         & -2.459                      & 0.0140                                 & *                    \\
Surgeon D : SurgeryType 11          & 0.1500                          & 0.0667                         & 2.249                       & 0.0246                                & *                    \\
Surgeon E : SurgeryType 1           & -0.4140                         & 0.1934                         & -2.14                       & 0.0325                                & *                    \\
Surgeon E : SurgeryType 2           & -0.3372                         & 0.1232                         & -2.737                      & 0.0062                                & **                   \\
Surgeon E : SurgeryType 7           & 0.4288                          & 0.1265                         & 3.389                       & 0.0007                                & ***                  \\
Surgeon F : SurgeryType 2           & -0.3618                         & 0.2367                         & -1.529                      & 0.1265                                &                      \\
Surgeon F : SurgeryType 3           & -0.2657                         & 0.1424                         & -1.866                      & 0.0622                                & .                    \\
\midrule[1pt]
log likelihood value               & -1040                           &                                &                             &                                         &                      \\
Adjusted $R^2$ value & 0.2276                          &                                &                             &                                         &\\
\bottomrule[1.5pt]
\end{tabular}
\end{table}
Given these results, it is meaningful to investigate the effects of the three interactions. However, there are too many variables in the models with interaction terms. We perform variable selection by applying the LASSO method.
In table \ref{tab:lassoModelV}, we report the coefficients relevant to the two predictor variables in Model V, \textit{i.e.}, $SURGEON$, $SURGERYTYPE$ and their interaction. The coefficients of other predictors (\textit{e.g.}, $WEEKDAY$ and $ORDER$) are not listed in the table, since they are quite similar to those of Model IV. Recall that Surgeon A is the baseline for $SURGEON$. Generally, the mean duration of surgeries performed by Surgeon B is lower than that of Surgeon A. Surgeon B needs more time to perform Surgery 2 and 11, but less time for Surgery 7. The mean of the duration of Surgeon C's surgeries is more than that of Surgeon A. Surgeon C is slower when he/she is performing Surgery 9. The mean durations of Surgeon D and E's surgeries are not significantly higher or lower than that of Surgeon A. However, Surgeon D spends less time on Surgery 5 and 9, but more time on Surgery 11. Surgeon E needs more time to perform Surgery 7, but less time to perform Surgery 1 and 2. Surgeon F is generally much faster than Surgeon A, especially when he/she is performing Surgery 3. Alternatively, from a surgery perspective, the reference is Surgery 4. It is illustrated that the mean of the duration of Surgery 1, 5, 6, and 9 is larger than that of Surgery 4, while that of Surgery 2, 8, 10, and 11 is smaller. In addition, Surgery 2 seems hard for Surgeon B, requiring more time, but easy for Surgeon E and F, and Surgery 11 seems hard for Surgeon B, C, and D.  

\begin{table}[h]
\centering
\caption{The regression results relevant to $SURGEON$, $ORDER$ and their interaction obtained by using Lasso for Model VI}
\label{tab:lassoModelVI}
\begin{tabular}{lrrrrl}
\toprule[1.5pt]
\multicolumn{1}{c}{Variable} & \multicolumn{1}{c}{Coefficient} & \multicolumn{1}{c}{Std. Error} & \multicolumn{1}{c}{t value} & \multicolumn{1}{c}{Pr($>|$t$|$)} & \multicolumn{1}{c}{} \\
\midrule[1pt]
Order 3$\sim$1                & -0.1547                         & 0.0319                         & -4.848                      & 1.33e-06                         & ***                  \\
Order 3$\sim$3                & -0.0847                         & 0.0311                         & -2.722                      & 0.0065                           & **                   \\
Order 4$\sim$1                & -0.1649                         & 0.0528                         & -3.123                      & 0.0018                           & **                   \\
Order 4$\sim$2                & -0.1018                         & 0.0555                         & -1.833                      & 0.0669                           & .                    \\
Order 4$\sim$4                & -0.1002                         & 0.0507                         & -1.975                      & 0.0484                           & *                    \\
Order 5$\sim$1                & -0.4219                         & 0.1696                         & -2.487                      & 0.0129                           & *                    \\
Surgeon C                     & 0.0743                          & 0.0222                         & 3.349                       & 0.0008                           & ***                  \\
Surgeon F                     & -0.2692                         & 0.0456                         & -5.908                      & 3.96e-09                         & ***                  \\
Surgeon B : Order 2$\sim$1    & -0.1406                         & 0.0393                         & -3.582                      & 0.0003                           & ***                  \\
Surgeon B : Order 2$\sim$2    & -0.1410                         & 0.0406                         & -3.471                      & 0.0005                           & ***                  \\
Surgeon B : Order 3$\sim$2    & -0.2284                         & 0.0701                         & -3.259                      & 0.0011                           & **                   \\
Surgeon B : Order 3$\sim$3    & -0.1713                         & 0.0752                         & -2.280                       & 0.0227                           & *                    \\
Surgeon B : Order 4$\sim$3    & -0.4338                         & 0.1063                         & -4.082                      & 4.62e-05                         & ***                  \\
Surgeon B : Order 5$\sim$5    & -0.3933                         & 0.2673                         & -1.471                      & 0.1414                           &                      \\
Surgeon C : Order 3$\sim$1    & 0.1501                          & 0.0713                         & 2.104                       & 0.0355                           & *                    \\
Surgeon C : Order 3$\sim$2    & -0.2581                         & 0.0687                         & -3.758                      & 0.0002                           & ***                  \\
Surgeon C : Order 5$\sim$4    & -0.7556                         & 0.2687                         & -2.812                      & 0.0050                            & **                   \\
Surgeon D : Order 5$\sim$1    & 0.9796                          & 0.4132                         & 2.371                       & 0.0178                           & *                    \\
Surgeon E : Order 3$\sim$2    & -0.3612                         & 0.2186                         & -1.652                      & 0.0987                           & .                    \\
Surgeon E : Order 3$\sim$3    & -0.4276                         & 0.1726                         & -2.478                      & 0.0133                           & *                    \\
\midrule[1pt]
log likelihood value          & -1055                     &                                &                             &                                  &                      \\
Adjusted $R^2$ value          & 0.2215                          &                                &                             &                                  &      \\
\bottomrule[1.5pt]               
\end{tabular}
\end{table}

In Table \ref{tab:lassoModelVI}, we report the coefficients of $SURGERON$, $ORDER$, and their interaction in Model VI. It is shown that Surgeon B and C are flexible. That is, the duration of these two surgeons' surgeries is significantly influenced by the surgery positions. In particular, Surgeon B (C) work much faster for the surgeries in position $4\sim 3$ ($5\sim 4$).
Surgeon C (D) performs surgeries in position $3\sim 1$ ($5\sim 1$) slowly. Surgeon E needs more time for the second and third surgeries when he/she has three surgeries in a day. In addition, the interaction between Surgeon F and surgery position is not significant. 

\begin{table}[h]
\centering
\caption{The regression results relevant to $ORDER$, $SURGERYTYPE$ and their interaction obtained by using Lasso for Model VII}
\label{tab:lassoModelVII}
\begin{tabular}{lrrrrl}
\toprule[1.5pt]
\multicolumn{1}{c}{Variable}      & \multicolumn{1}{c}{Coefficient} & \multicolumn{1}{c}{Std. Error} & \multicolumn{1}{c}{t value} & \multicolumn{1}{c}{Pr($>|$t$|$)} &     \\
\midrule[1pt]
Order 3$\sim$1                     & -0.1245                         & 0.0330                         & -3.776                      & 0.0002                                  & *** \\
Order 3$\sim$2                     & -0.0759                         & 0.0281                         & -2.700                      & 0.0070                                  & **  \\
Order 4$\sim$1                     & -0.0770                         & 0.0548                         & -1.406                      & 0.1599                                  &     \\
Order 4$\sim$4                     & -0.0903                         & 0.0509                         & -1.773                      & 0.0763                                  & .   \\
SurgeryType 1                      & 0.1651                          & 0.0392                         & 4.214                       & 2.60e-05                                & *** \\
SurgeryType 2                      & -0.3132                         & 0.0521                         & -6.008                      & 2.17e-09                                & *** \\
SurgeryType 5                      & 0.4212                          & 0.0753                         & 5.591                       & 2.51e-08                                & *** \\
SurgeryType 6                      & 0.1262                          & 0.0254                         & 4.961                       & 7.51e-07                                & *** \\
SurgeryType 9                      & 0.7586                          & 0.0558                         & 13.594                      & \textless2.00e-16                       & *** \\
SurgeryType 10                     & -0.1614                         & 0.0335                         & -4.823                      & 1.51e-06                                & *** \\
SurgeryType 2 : Order 3$\sim$3     & -0.6483                         & 0.1960                         & -3.308                      & 0.0010                                  & *** \\
SurgeryType 3 : Order 3$\sim$1     & 0.1381                          & 0.0646                         & 2.137                       & 0.0327                                  & *   \\
SurgeryType 3 : Order 5$\sim$4     & -0.6374                         & 0.2677                         & -2.381                      & 0.0174                                  & *   \\
SurgeryType 5 : Order 3$\sim$2     & -0.8241                         & 0.3849                         & -2.141                      & 0.0324                                  & *   \\
SurgeryType 6 : Order 3$\sim$1     & -0.2692                         & 0.1319                         & -2.041                      & 0.0414                                  & *   \\
SurgeryType 6 : Order 4$\sim$1     & -0.4893                         & 0.2252                         & -2.173                      & 0.0299                                  & *   \\
SurgeryType 7 : Order 3$\sim$1     & 0.4146                          & 0.2682                         & 1.546                       & 0.1223                                  &     \\
SurgeryType 7 : Order 3$\sim$3     & -0.5595                         & 0.2664                         & -2.100                      & 0.0358                                  & *   \\
SurgeryType 7 : Order 4$\sim$4     & 0.5464                          & 0.3797                         & 1.439                       & 0.1503                                  &     \\
SurgeryType 8 : Order 3$\sim$3     & -0.4080                         & 0.1887                         & -2.162                      & 0.0307                                  & *   \\
SurgeryType 9 : Order 4$\sim$4     & 0.8681                          & 0.3842                         & 2.260                       & 0.0239                                  & *   \\
SurgeryType 10 : Order 4$\sim$1    & -0.4181                         & 0.2741                         & -1.526                      & 0.1272                                  &     \\
SurgeryType 11 : Order 3$\sim$3    & -0.3565                         & 0.0761                         & -4.685                      & 2.95e-06                                & *** \\
SurgeryType 11 : Order 5$\sim$1    & -1.4652                         & 0.3764                         & -3.892                      & 0.0001                                  & *** \\
\midrule[1pt]
log likelihood value               & -1050                           &                                &                             &                                         &     \\
Adjusted $R^2$ value & 0.224                           &                                &                             &                                         &    \\
\midrule[1.5pt]
\end{tabular}
\end{table}

We report the coefficients relevant to $SURGERTYPE$, $ORDER$, and their interaction in Table \ref{tab:lassoModelVII}. As we know, surgery duration greatly depends on surgery type. However, this table shows that the effects of surgery type are different for surgeries with different surgery positions.  
For example, the duration of Surgery 2 decreases if it is the last surgery in a day with three surgeries performed by a surgeon that day. A similar interpretation can be done for all the interaction items. It is worth noting that Surgery 3 is flexible, since its duration increases if it is placed in the position $3\sim 1$, while it decreases a lot if it is in the position $5\sim 4$. 

In sum, the results of the models with interactions support Hypothesis 5, \textit{i.e.}, the interaction between some factors influence surgery duration. 
 The log likelihood and adjusted $R^2$ values of the models with interactions are larger than those of simple models, which indicates the necessity to consider these interactions.
We find that surgeons exhibit different performance on different surgery types, as well as on surgeries in different positions. Also, the interaction between surgery type and surgery position impacts the duration.

\section{Conclusions and Future Work}
In this paper, we investigate the impact of clinical and
non-clinical factors on surgery duration. It is found that surgery duration is influenced by surgeons' workload and workload in the OR where the surgery is performed. The duration decreases with surgeons' workload. However, it does not monotonically decrease with the workload in the OR. It decreases with the OR workload if the number of surgeries in the OR in a day is not more than four, while it increases with the OR workload if the number is beyond four. Also, we find the duration of a surgery is slightly impacted by the position of the surgery in a sequence of surgeries a surgeon performs in a day. In addition, different surgeons exhibit different patterns on the impacts of surgery types and surgery positions. More specifically, some surgeons perform certain surgery types faster than others. Surgeon B and C are flexible, whose performance is largely dependent on surgery positions. Furthermore, it is found that the effect of interactions between surgery types and positions is significant. In other words, for some kinds of surgeries, the duration is different if it is placed in different positions. These results all indicate reason to incorporate non-clinical factors in OR planning and scheduling. More accurate schedules may be attained, with fewer delays and less re-scheduling as a result.

Note that the surgeon performance of surgery duration is not equivalent to the quality of surgery from a medical perspective. Hence, an intuitive idea on future work is to investigate the impacts of the factors discussed in this paper on healthcare quality, \textit{e.g.}, readmission and mortality rates. Also, the impart of surgery duration on healthcare quality should be take into consideration. Furthermore, as mentioned before, we only focus on the third part of the whole surgery process, \textit{i.e.}, surgery duration. Hence, another future work could be to investigate the impacts of the factors mentioned in this paper on the other time intervals, as well as the impacts of the length of the intervals on healthcare quality. 
\section*{Acknowledgments}
This work was supported in part by Research Grants Council (RGC) Theme-Based Research Scheme under Grant T32-102/14-N, and in part by the National Natural Science Foundation of China (NSFC) under Grant 71701132 and Grant 61603321.
\bibliographystyle{ormsv080}


\end{document}